\documentstyle[prb,aps,floats,epsfig]{revtex}%%prp%%
\begin{document}
%-------------------------------------------------------------------------------
%one-column-prb-style for the title and abstract--> goc-to-prb or prb-to-goc
\twocolumn[\hsize\textwidth\columnwidth\hsize\csname@twocolumnfalse%%prb%%
\endcsname%%prb%%
%-------------------------------------------------------------------------------
\title{Integrability and level crossing manifolds \\ in a quantum Hamiltonian
  system}  

\author{Vyacheslav V. Stepanov and Gerhard M\"uller} 

\address{Department of Physics, University of Rhode Island, Kingston RI
  02881-0817} 

\date{\today~--~2.5}
\maketitle
\begin{abstract}
  We consider a two-spin model, represented {\it classically} by a nonlinear
  autonomous Hamiltonian system with two degrees of freedom and a nontrivial
  integrability condition, and {\it quantum mechanically} by a real symmetric
  Hamiltonian matrix with invariant blocks of dimensionalities $K = \frac{1}{2}
  l(l+1),~l=1, 2, \ldots$ In the six-dimensional parameter space of this model,
  classical integrability is satisfied on a five-dimensional hypersurface, and
  level crossings occur on four-dimensional manifolds that are completely
  embedded in the integrability hypersurface except for some lower-dimensional
  sub-manifolds. Under mild assumptions, the classical integrability condition
  can be reconstructed from a purely quantum mechanical study of level
  degeneracies in finite-dimensional invariant blocks of the Hamiltonian matrix.
  Our conclusions are based on rigorous results for $K=3$ and on numerical
  results for $K=6,10$.
\end{abstract}
%\pagebreak
%insert suggested PACS number in the next line
%\pagebreak
\twocolumn%%prb%%
%-------------------------------------------------------------------------------
%end of one column-prb-style
]%%prb%%
%-------------------------------------------------------------------------------
%gerhard-one-column-draft style, centered
%%goc%%\newpage
%%goc%%\narrowtext
%%goc%%\oddsidemargin=4cm
%%prp%%\onecolumn
%%%%%%%%%%%%%%%%%%%%%%%%%%%%%%%%%%%%%%%%%%%%%%%%%%%%%%%%%%%%%%%%%%%%%%%%%%%%%%%%
%
\section{Introduction}\label{secI}
%
%%%%%%%%%%%%%%%%%%%%%%%%%%%%%%%%%%%%%%%%%%%%%%%%%%%%%%%%%%%%%%%%%%%%%%%%%%%%%%%%
One of the most widely studied indicators of quantum chaos can be obtained via
the statistical analysis of energy level spacings. Generically, the level
spacings of quantized integrable systems tend to be well described by an
exponential distribution (Poisson statistics), whereas quantized nonintegrable
systems tend to have a distribution in which the probability of very
small spacings is suppressed (Wigner statistics) due to the phenomenon of level
repulsion. The level turbulence such as exists in quantized nonintegrable
systems can be simulated by the eigenvalues of random matrices with specific
distributions of elements (e.g. Gaussian orthogonal ensemble).\cite{Gut90,Rei92}

The statistical nature of this indicator precludes its use for mapping out the
regions of integrability in the parameter space of Hamiltonian systems. However,
determining the conditions for the occurrence of level degeneracies, on which
the outcome of the statistical analysis depends, proves to be useful for
precisley that purpose.

Here we show for a specific model system how the (known) classical
integrability condition in a six-dimensional (6D) parameter space can be
reconstructed, under mild assumptions, from a purely quantum mechanical study of
the manifolds (in the same parameter space) where at least two energy levels are
degenerate. 

Practical considerations dictate that we use a model system where the Hilbert
space splits into finite-dimensional invariant subspaces. However, the
significance of the results presented here transcend this restriction and
suggest that the concept of integrability remains meaningful albeit more subtle
for quantum systems with few degrees of freedom.\cite{Wei92,WM95}

We consider two quantum spins ${\bf S}_1$, ${\bf S}_2$ in biaxial orientational
potentials interacting via a biaxial exchange coupling. The Hamiltonian reads
\begin{equation}
H=\sum_{\alpha=xyz} \left\{ -J_{\alpha}S^{\alpha}_{1} S^{\alpha}_{2}+
\frac 12 A_{\alpha}
\left[ (S^{\alpha}_{1})^{2} +(S^{\alpha}_{2})^{2} \right] \right\}.
\label{Ham}
\end{equation}
The spin operators ${\bf S}_l=\left( S^x_l, S^y_l, S^z_l \right)$ satisfy the
commutation relations $[ S^\alpha_l,S^\beta_{l'} ]= i\hbar\delta_{ll'}
\sum_\gamma \epsilon_{\alpha \beta \gamma} S^\gamma_l$. Their time evolution is
governed by the Heisenberg equation
\begin{equation}
\frac{d{\bf S}_l}{dt}=\frac i\hbar [ H, {\bf S}_l ], ~~~ l=1,2.
\label{Heis}
\end{equation}

If both spins have the same quantum mechanical length $\sqrt{\sigma(\sigma+1)}$
($\sigma=\frac{1}{2},1,\frac{3}{2},\dots$), the discrete symmetry group of the
Hamiltonian~(\ref{Ham}) is $D_2 \otimes S_2$, where $D_2$ contains all the
twofold rotations $C^\alpha_2$, $\alpha=x,y,z$ about the coordinate axes, and
$S=(E,P)$ is the permutation group of the two spins. The characters of this
group are displayed in Table~\ref{tab1}.\cite{Web88}

%%%%%%%%%%%%%%%%%%%%%%%%%%%%%%%%BEGIN-TABLE%%%%%%%%%%%%%%%%%%%%%%%%%%%%%%%%%%%%%%
\begin{table}[ht]
\begin{tabular}{c|rrrrrrrr}
$D_{2} \otimes S_{2}$ & $E$ & $C_{2}^{z}$ & $C_{2}^{y}$ & $C_{2}^{x}$ &
$P$ & $PC_{2}^{z}$
 & $PC_{2}^{y}$ & $PC_{2}^{x}$ \\ \hline
A1S & 1 & 1 & 1 & 1 & 1 & 1 & 1 & 1 \\
A1A & 1 & 1 & 1 & 1 & -1 & -1 & -1 & -1 \\
B1S & 1 & 1 & -1 & -1 & 1 & 1 & -1 & -1 \\
B1A & 1 & 1 & -1 & -1 & -1 & -1 & 1 & 1 \\
B2S & 1 & -1 & 1 & -1 & 1 & -1 & 1 & -1 \\
B2A & 1 & -1 & 1 & -1 & -1 & 1 & -1 & 1 \\
B3S & 1 & -1 & -1 & 1 & 1 & -1 & -1 & 1 \\ 
B3A & 1 & -1 & -1 & 1 & -1 & 1 & 1 & -1 \\ 
\end{tabular}
\caption[T1]{The characters of the irreducible representations $R$ of the group
$D_{2} \otimes S_{2}$.}
\label{tab1}
\end{table}
%%%%%%%%%%%%%%%%%%%%%%%%%%%%%%END-TABLE%%%%%%%%%%%%%%%%%%%%%%%%%%%%%%%%%%%%%%%%%%

The use of symmetry-adapted basis vectors with transformation properties
corresponding to the eight different irreducible representations $R$ of $D_{2}
\otimes S_{2}$ brings the Hamiltonian matrix into block-diagonal form:
\begin{equation}\label{HRsigma}
H = \bigoplus_{R,\sigma} H_R^\sigma.
\end{equation}
There exist invariant subspaces with dimensionalities $K = 1, 3, 6, 10, \ldots$
in 16 different realizations for four different values of the spin quantum
number $\sigma$ as illustrated in Table~\ref{tab2}. The case $K=1$ is
exceptional. 

%%%%%%%%%%%%%%%%%%%%%%%%%%%BEGIN-TABLE%%%%%%%%%%%%%%%%%%%%%%%%%%%%%%%%%%%%%%%%%%
\begin{table}[ht]
\begin{tabular}{c|rrrrrrrr}
$R\backslash\sigma$ & $\frac 12$ & 1 & $\frac 32$ & 2 & $\frac 52$ & 3 & 
$\frac72$ & 4 \\ 
\hline
$A1S$ & -- & {\bf 3} & 1 & 6 & {\bf 3} & 10 & 6 & 15 \\ 
$A1A$ & 1 & -- & {\bf 3} & 1 & 6 & {\bf 3} & 10 & 6 \\ 
$B1S$ & 1 & 1 & {\bf 3} & {\bf 3} & 6 & 6 & 10 & 10 \\ 
$B1A$ & -- & 1 & 1 & {\bf 3} & {\bf 3} & 6 & 6 & 10 \\ 
$B2S$ & 1 & 1 & {\bf 3} & {\bf 3} & 6 & 6 & 10 & 10 \\ 
$B2A$ & -- & 1 & 1 & {\bf 3} & {\bf 3} & 6 & 6 & 10 \\ 
$B3S$ & 1 & 1 & {\bf 3} & {\bf 3} & 6 & 6 & 10 & 10 \\ 
$B3A$ & -- & 1 & 1 & {\bf 3} & {\bf 3} & 6 & 6 & 10 \\ 
\end{tabular}
\caption[T2]{Dimensionalities $K$ of the invariant subspaces pertaining to the
  eight symmetry classes $R$ of eigenstates for spin quantum numbers
  $\sigma\leq 4$.}  
\label{tab2}
\end{table}
%%%%%%%%%%%%%%%%%%%%%%%%%%%%%END-TABLE%%%%%%%%%%%%%%%%%%%%%%%%%%%%%%%%%%%%%%%%%%

%%%%%%%%%%%%%%%%%%%%%%%%%%%%%%%%%%%%%%%%%%%%%%%%%%%%%%%%%%%%%%%%%%%%%%%%%%%%%%%%
%
\section{Classical integrability manifold}\label{secII}
%
%%%%%%%%%%%%%%%%%%%%%%%%%%%%%%%%%%%%%%%%%%%%%%%%%%%%%%%%%%%%%%%%%%%%%%%%%%%%%%%%
In the limit $\hbar \rightarrow 0$, $\sigma \rightarrow \infty$,
$\hbar\sqrt{\sigma(\sigma+1)} = s$, the operators $S^\alpha_l$ become the
components of the classical spin vector with fixed length $s$,
\begin{eqnarray}
{\bf S}_l &=& \left( S^x_l, S^y_l, S^z_l \right) \nonumber \\
 &=& s\left(\sin\vartheta_l \cos\varphi_l,
\sin\vartheta_l \sin\varphi_l, \cos\vartheta_l \right),
\label{Spin}
\end{eqnarray}
and (\ref{Heis}) turns into Hamilton's equation,
\begin{equation}
\frac{d{\bf S}_l}{dt}=- {\bf S}_l \times \frac{\partial H}{\partial {\bf S}_l}=
\left\{ H, {\bf S}_l \right\},~~ l=1,2~,
\end{equation}
where $\{ S^\alpha_l, S^\beta_{l'} \}=-\delta_{ll'} \sum_\gamma
\epsilon_{\alpha\beta \gamma} S^\gamma_l$ are the Poisson brackets for spin
variables.  Each classical spin~(\ref{Spin}) is expressible in terms of two
canonical coordinates
\begin{equation}
p_l= s \cos\vartheta_l,~~q_l=\varphi_l,~~l=1,2.
\end{equation}

The Hamiltonian~(\ref{Ham}), now interpreted as a classical energy function,
thus specifies an autonomous system with two degrees of freedom. Integrability
of that system requires the existence of a second integral of the motion,
i.e. an analytic function $I$ of the spin components $S_l^\alpha$ with the
property $\{I,H\}=0$. 

A systematic search for a second invariant in the form of a degree-two
polynomial yielded two distinct nontrivial solutions, provided the six
parameters satisfy the condition\cite{MTWKM87} 
\begin{eqnarray}
 &&(A_{x}-A_{y})(A_{y}-A_{z})(A_{z}-A_{x}) \nonumber \\ 
&&\hspace*{2cm}+ \sum_{\alpha \beta \gamma = {\rm cycl}(xyz)} 
J^{2}_{\alpha}(A_{\beta}-A_{\gamma}) = 0.
\label{Int}
\end{eqnarray}
If there is no single-site anisotropy, $A_x=A_y=A_z$, then the second integral
of motion reads
\begin{eqnarray}
I = &-&\sum_{\alpha\beta\gamma=cycl(xyz)} J_{\alpha} J_{\beta}
S^{\gamma}_{1} S^{\gamma}_{2} \nonumber \\ &+&
\frac 12 \sum_{\alpha=xyz} J_{\alpha}^2 \left[ \left( S_1^{\alpha} \right)^2 + 
\left( S_2^{\alpha} \right)^2 \right],
\label{Op1}
\end{eqnarray}
otherwise it has the form
\begin{eqnarray}\label{Op2}
&&\hspace*{1.8cm} I = \sum_{\alpha=xyz} g_{\alpha} S^{\alpha}_{1} S^{\alpha}_{2},
\\
g_\alpha &=& J_\alpha ( J_\alpha+J_\beta+J_\gamma )+
(A_{\alpha}-A_{\beta}) J_\gamma + ( A_\alpha-A_\gamma ) J_\beta \nonumber \\
&-& \left( A_\alpha - A_\beta \right) \left( A_\alpha-A_\gamma \right),~~
\alpha\beta\gamma={\rm cycl}(xyz). \nonumber
\end{eqnarray}

Hence, in the 6D parameter space of this two-spin model the classical
integrability condition is satisfied on a 5D manifold. Integrals of the motion
of higher-degree polynomial form or of non-polynomial form cannot be ruled out,
but it is unlikely that any other hypersurface of integrability would have
escaped the numerical studies of this model.\cite{MTWKM87,SKMWT88} Additional
integrability manifolds of dimensionalities four or less remain an intriguing
possibility but do not interfere with any conclusions reached in this study.

%%%%%%%%%%%%%%%%%%%%%%%%%%%%%%%%%%%%%%%%%%%%%%%%%%%%%%%%%%%%%%%%%%%%%%%%%%%%%%%%
%
\section{Level crossing manifolds}\label{secIII}
%
%%%%%%%%%%%%%%%%%%%%%%%%%%%%%%%%%%%%%%%%%%%%%%%%%%%%%%%%%%%%%%%%%%%%%%%%%%%%%%%%
Does the integrability condition (\ref{Int}) of the classical two-spin model
(\ref{Ham}) have any bearing on the presence or absence of level degeneracies in
low-dimensional invariant subspaces of the corresponding quantum two-spin model?
The subspaces with a single energy level $(K=1)$, which are realized for $\sigma
\leq 2$, are uninteresting in this context. The next
lowest subspace dimensionality is $K=3$.  The occurrence of level degeneracies
for the parametric Hamiltonian (\ref{Ham}) will now be analyzed on a rigorous
basis for all 16 invariant subspaces with $K=3$. Their entries are highlighted
in Table~\ref{tab2}.

%%%%%%%%%%%%%%%%%%%%%%%%%%%%%%%%%%%%%%%%%%%%%%%%%%%%%%%%%%%%%%%%%%%%%%%%%%%%%%%%
%
\subsection{Parametric representation for $K=3$}\label{secIIIa}
%
%%%%%%%%%%%%%%%%%%%%%%%%%%%%%%%%%%%%%%%%%%%%%%%%%%%%%%%%%%%%%%%%%%%%%%%%%%%%%%%%
The most general real symmetric $3\times 3$ matrix has six independent elements.
For the purpose of studying level degeneracies, it is sufficient to consider
matrices with zero trace:
\begin{equation}
M = \left( 
\begin{array}{ccc}
2h & b & d \\
b & e-h & c \\
d & c & -e-h 
\end{array}
\right).
\label{matrix}
\end{equation}
That leaves five independent elements $b$, $c$, $d$, $e$, $h$ and thus
simplifies the analysis because the characteristic polynomial now has a
vanishing quadratic term:
\begin{equation}\label{charpol}
\left| M - xE \right| = x^3 - Bx + C = 0 .
\end{equation}
The discriminant has the form
\begin{equation}
D = 4B^3 - 27C^2
\end{equation}
with coefficients
\begin{eqnarray}
B & = & b^2 + c^2 + d^2 + e^2 + 3h^2, \\
C & = & h \left( 2e^2 + 2c^2 - b^2 - d^2 - 2h^2 \right) + e \left( d^2
- b^2 \right) -2 bcd. \nonumber
\end{eqnarray}
The zeros of $D$ coincide with the points of level degeneracy in $M$. This
is evident in the product form
\begin{equation}
D = \prod_{i < k} (x_i - x_k)^2
\end{equation}
of the discriminant in terms of the roots $x_i$ of (\ref{charpol}). Since $D$ is
non-negative and depends smoothly on $b$, $c$, $d$, $e$, $h$, its partial
derivatives must also vanish at all points of level degeneracy:
\begin{eqnarray}\label{fiverel}
\frac{\partial D}{\partial b} & = & 12 B^2 2b - 54 C \left( -2bh - 2eb
-2cd \right) = 0, \nonumber \\
\frac{\partial D}{\partial c} & = & 12 B^2 2c - 54 C \left( 4hc - 2bd \right) =
0, \nonumber \\ 
\frac{\partial D}{\partial d} & = & 12 B^2 2d - 54 C \left( -2dh +2ed
-2cb \right) = 0, \\
\frac{\partial D}{\partial e} & = & 12 B^2 2e - 54 C \left( 4eh + d^2 -
b^2 \right) = 0, \nonumber \\ 
\frac{\partial D}{\partial h} & = & 12 B^2 6h - 
                                 54 C \left( 2e^2 + 2c^2 - b^2 - d^2 - 6h^2
                                 \right) = 0. \nonumber
\end{eqnarray}
These additional conditions simplify the search for zeros of $D$. $C=0$ implies
$B=0$ and vice versa. This case describes the threefold level degeneracy at
$b=c=d=e=h=0$. Henceforth we assume $B \neq 0$ and $C \neq 0$ with no loss of
generality.  The five relations~(\ref{fiverel}) can then be written in the more
compact form
\begin{eqnarray}
\frac{2 B^2}{9 C} & = & \frac{-bh-be-cd}{b} = \frac{2hc -bd}{c} =
\frac{-dh+ed-bc}{d} \nonumber \\
 & = & \frac{4eh+d^2-b^2}{2e} = \frac{2e^2+2c^2-b^2-d^2 -6h^2}{6h}.
\label{fiverel2}
\end{eqnarray}

Inspection shows that only two of the relations (\ref{fiverel2}) are
independent, and that $D=0$ holds wherever (\ref{fiverel2}) is satisfied. The
points of level crossing are thus confined to a 3D manifold in
$(b,c,d,e,h)$-space. This manifold can be parametrized by three of the five
elements. For $e \neq 0$ and $b\neq \pm d$ we have
\begin{equation}\label{lcman5}
c=\frac{2bde}{b^2-d^2}\,,\;\;\;
h=\frac{b^2-d^2}{6e}\left[1-\frac{2e^2(b^2+d^2)}{(b^2-d^2)^2}\right]\,.
\label{genpar}
\end{equation}

Viewed on any of the five 4D coordinate hyperplanes, where one of the elements
$b, c, d, e, h$ is equal to zero, the level crossing manifold reduces to two or
three 2D surfaces. Parametric representations of all eight such surfaces are
given in Table~\ref{tab3}.
%%%%%%%%%%%%%%%%%%%%%%%%%%%%%%BEGIN-TABLE%%%%%%%%%%%%%%%%%%%%%%%%%%%%%%%%%%%%%%%
\begin{table}[htb]
\begin{tabular}{rcrcc}
 $b$ & $c$ & $d$ & $e$ & $h$ \\ \hline 
 0 & 0 & * & * & ${\displaystyle\frac{2e^2 - d^2}{6e}}$ \\ 
 0 & * & 0 & * & ${\displaystyle \pm \frac 13 \sqrt{e^2 + c^2}}$ \\
 $\ast$ & 0 & 0 & * & ${\displaystyle\frac{b^2 - 2e^2}{6e}}$ \\ 
 $\ast$ & * & $\pm b$ & 0 & ${\displaystyle \pm \frac{b^2 - c^2}{3c}}$ \\
 $\ast$ & ${\displaystyle \pm \frac{\sqrt{2} bd}{\sqrt{b^2 + d^2}}}$ & * 
 & ${\displaystyle \pm \frac{1}{\sqrt{2}} \frac{b^2 - d^2}{\sqrt{b^2 +d^2}}}$ 
 & 0 \\
\end{tabular}
\caption{2D intersections of the 3D level crossing manifold (\ref{lcman5}) with
 the 4D coordinate hyperplanes. The two elements of (\ref{matrix}) which play
the role of parameters in each case are marked by asterisks. }
\label{tab3}
\end{table}
%%%%%%%%%%%%%%%%%%%%%%%%%%%%%END-TABLE%%%%%%%%%%%%%%%%%%%%%%%%%%%%%%%%%%%%%%%%%%

%%%%%%%%%%%%%%%%%%%%%%%%%%%%%%%%%%%%%%%%%%%%%%%%%%%%%%%%%%%%%%%%%%%%%%%%%%%%%%%%
%
\subsection{Level crossing labels}\label{secIIIb}
%
%%%%%%%%%%%%%%%%%%%%%%%%%%%%%%%%%%%%%%%%%%%%%%%%%%%%%%%%%%%%%%%%%%%%%%%%%%%%%%%%
In a three-level system, any twofold degeneracy either involves the upper two
levels or the lower two levels. How does this distinction manifest itself in the
structure of the level crossing manifold? The eigenvalues of the matrix $M$ for
points on the level crossing manifold can be written in the form
\begin{equation}\label{x123xi}
(x_1, x_2, x_3) = \left(\xi, -\frac{1}{2}\xi, -\frac{1}{2}\xi\right), 
\end{equation}
where
\begin{equation}\label{xi}
\xi = \frac{4}{3}\left[\frac{e(b^2+d^2)}{b^2-d^2} + \frac{b^2-d^2}{4e}\right].
\end{equation}
If $\xi>0$ ($\xi<0$) then it is the highest (lowest) level that remains
non-degenerate. A threefold degeneracy $(\xi=0)$ occurs only at the point
$b=c=d=e=h=0$. 

Do the points with $\xi>0$ and the points with $\xi<0$ form connected regions on
the level-crossing manifold? To investigate this issue we consider the map
described by (\ref{lcman5}) between the $(b,d,e)$-space and the 3D level
crossing manifold in $(b,c,d,e,h)$-space. This map is singular on the three
planes $e=0$, $b+d=0$, $b-d=0$, which divide the $(b,d,e)$-space into octants.
Octants which share a face (one quadrant of a coordinate plane) have
$\xi$-values of opposite sign, and octants which share only an edge (half a
coordinate axis) have $\xi$-values of equal sign.

For a point $(b,d,e)$ approaching any one of the three planes that separate
octants, the image in $(b,c,d,e,h)$-space diverges, but for a point
$(b,d,e)$ approaching a line where any two of the three separating planes
intersect, the image may or may not diverge.

Consider smooth trajectories of points $(b,d,e)$ that connect two octants across
one of these special lines. Inspection shows that any
trajectory connecting octants with a common face has a divergent image.
However, there do exist trajectories with non-divergent and continuous images
between any two octants that have only an edge in common.

For example, set $b+d>0$ and consider trajectories $e\to0$, $b-d\to0$ with
$e/(b-d)=u\neq 0$ toward the edge of four octants. Along such a trajectory we
have
\begin{equation}\label{octpath}
c=bu,~~ h=\frac{b}{3u}-\frac{bu}{3},~~ \xi=\frac{4bu}{3}+\frac{4b}{6u}.
\end{equation}
Octants that are diagonally across the edge have either both $u>0$ or $u<0$.
Hence they are connected by trajectories with finite $c,h$ and with no change of
sign in $\xi$. No such trajectories exist between adjacent octants.

All this demonstrates that the 3D level crossing manifold consists of one sheet
for $\xi<0$ and one sheet for $\xi>0$, connected only at the point
$b=c=d=e=h=0$.

%%%%%%%%%%%%%%%%%%%%%%%%%%%%%%%%%%%%%%%%%%%%%%%%%%%%%%%%%%%%%%%%%%%%%%%%%%%%%%%%
%
\subsection{Embedment in classical integrability manifold}\label{secIIIc} 
%
%%%%%%%%%%%%%%%%%%%%%%%%%%%%%%%%%%%%%%%%%%%%%%%%%%%%%%%%%%%%%%%%%%%%%%%%%%%%%%%%

These results can now be used to locate all level crossings in the invariant
blocks of (\ref{HRsigma}) with $K=3$. Table~\ref{tab2} identifies 16 such
blocks, two for each symmetry class. The three eigenvalues of $H_R^\sigma$ on
the level crossing manifold are then
\begin{equation}
(E_1, E_2, E_3) =  \left(\xi, -\frac{1}{2}\xi, -\frac{1}{2}\xi\right) 
+ \lambda_R^\sigma
\end{equation}
with $\xi$ given in (\ref{xi}). Table~\ref{tab4} expresses $\lambda_R^\sigma =
{\mathrm Tr}H_R^\sigma$ and the matrix elements $a,b,c,d,e,h$ in terms of the six
Hamiltonian parameters for four of the 16 invariant subspaces of (\ref{HRsigma})
with $K=3$.

%%%%%%%%%%%%%%%%%%%%%%%%%%%%%%BEGIN-TABLE%%%%%%%%%%%%%%%%%%%%%%%%%%%%%%%%%%%%%%%
\begin{table}[htb]
\begin{tabular}{l|l}
& ~~$\lambda = 2(A_x+A_y+A_z)/3$ \\
$H_{A1S}^1$~~ &  ~~$\displaystyle b=-(J_x+J_y)/\sqrt{2}$~~~
 $\displaystyle c=( A_x-A_y)/2$ \\
 &
 ~~$\displaystyle d=( J_y-J_x)/\sqrt{2}$~~~
 $\displaystyle e=-J_z$ \\
 &
 ~~$\displaystyle h=(A_x+A_y)/6 - A_z/3$
 \\ \hline
& ~~$\lambda = 4(A_x+A_y+A_z)$ \\
$H_{A1A}^3$~~ &  ~~$\displaystyle b=-3(J_x+J_y)/\sqrt{2}$~~~
 $\displaystyle c=3( A_x-A_y)/2$ \\
 &
 ~~$\displaystyle d=3( J_y-J_x )/\sqrt{2}$~~~
 $\displaystyle e=3J_z$ \\
 &
 ~~$\displaystyle h=( A_x+A_y)/2-A_z$ \\
\hline
& ~~$\lambda = 11(A_x+A_y)/6 + 7A_z/3 - 5J_z/3$ \\
$H_{B1S}^2$~~ & ~~$\displaystyle b=-\sqrt{3}( J_x+J_y)$~~~
 $\displaystyle c=\sqrt{3}( A_x-A_y)/2$ \\
 &
 ~~$\displaystyle d=J_y-J_x$~~~
 $\displaystyle e=(A_x+A_y)/2 - A_z + 2 J_z$ \\
 &
 ~~$\displaystyle h=(A_x+A_y)/3 - 2A_z/3 + J_z/3$ \\
\hline
& ~~$\lambda = 11(A_x+A_y)/6 + 7A_z/3 + 5J_z/3$ \\
$H_{B1A}^2$~~ & ~~$\displaystyle b=\sqrt{3}(J_y-J_x)$~~~
 $\displaystyle c=\sqrt{3}( A_x-A_y)/2$ \\
 &
 ~~$\displaystyle d=-J_x-J_y$~~~
 $\displaystyle e=(A_x+A_y)/2 - A_z - 2 J_z$ \\
 &
 ~~$\displaystyle h=(A_x+A_y)/3 - 2A_z/3 - J_z/3$ \\
\end{tabular}
\caption[]{Dependence of $\lambda_R^\sigma = {\mathrm Tr}H_R^\sigma$ and of the
  five independent matrix elements of $M = H_R^\sigma - \lambda_R^\sigma$ on the
  six parameters of (\ref{Ham}) for four invariant blocks of (\ref{HRsigma}) with
  $K=3$.}
\label{tab4}
\end{table}
%%%%%%%%%%%%%%%%%%%%%%%%%%%%%END-TABLE%%%%%%%%%%%%%%%%%%%%%%%%%%%%%%%%%%%%%%%%%%

Consider, for example, the matrix $H_{A1A}^3$ pertaining to the symmetry class
$A1A$ for spin quantum number $\sigma = 3$. If we take one of the relations
(\ref{fiverel2}) which must be satisfied at all points of level crossing,
\begin{equation}
c \left( 2e^2+2c^2-b^2-d^2-6h^2 \right) = 6h \left( 2ch - bd \right),
\end{equation}
and express the matrix elements in terms of the Hamiltonian parameters for
$H_{A1A}^3$, we find that it is equivalent to the classical integrability
condition~(\ref{Int})!  Hence no level crossings occur in $H_{A1A}^3$ if the
classical system is nonintegrable. In the 6D parameter space of (\ref{Ham}), the
points of level degeneracy pertaining to $H_{A1A}^3$ are thus confined to a 4D
manifold which is determined, according to (\ref{genpar}), by the two relations
\begin{eqnarray}
A_x-A_y &=& \frac{(J_x^2-J_y^2)J_z^2}{J_xJ_yJ_z}, \nonumber \\
A_x+A_y-2A_z &=& \frac{2J_x^2J_y^2-(J_x^2+J_y^2)J_z^2}{J_xJ_yJ_z}.
\end{eqnarray}
Either relation can be replaced by the classical integrability condition
(\ref{Int}).

We have determined that in all 16 invariant subspaces with $K=3$ the conditions
(\ref{genpar}) for the occurrence of a level degeneracy imply that the classical
integrability condition (\ref{Int}) is satisfied. Geometrically speaking, the
classical integrability condition is satisfied on a 5D hypersurface in 6D
parameter space. In each of the 16 invariant $K=3$ subspaces of $H$, level
crossings occur on a distinct 4D manifold. The result of our calculation is that
all 16 4D {\it level-crossing manifolds} are embedded in the 5D {\it classical
  integrability hypersurface} of the 6D parameter space.

%%%%%%%%%%%%%%%%%%%%%%%%%%%%%%%%%%%%%%%%%%%%%%%%%%%%%%%%%%%%%%%%%%%%%%%%%%%%%%%%
%
\subsection{Shape for $K=3$}\label{secIIId}
\begin{figure}[ht]
\centerline{\hspace{1mm}\epsfig{file=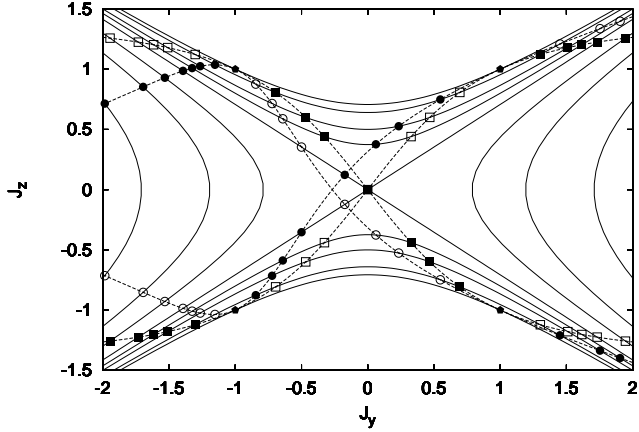,width=6.0cm,angle=-90}}
\caption{The dashed curves are level crossing lines in the reduced parameter
  space $(J_y,J_z,A)$ projected onto the $(J_y,J_z)$-plane for two invariant
  blocks $H_R^\sigma$ with $K=3$: $H_{A1A}^3$ (circles) and $H_{B1S}^2$
  (squares). The solid lines represent the integrability hyperboloid at $|A|=0.,
  0.3, 0.5, 0.6, 0.7, 1/\sqrt{2}, 0.9, 1.1, 1.4$. The pentagons mark symmetry
  points of $H$.}
\label{fig1}
\end{figure}

%
%%%%%%%%%%%%%%%%%%%%%%%%%%%%%%%%%%%%%%%%%%%%%%%%%%%%%%%%%%%%%%%%%%%%%%%%%%%%%%%%
For a graphical representation of the level crossing manifolds embedded in the
integrability manifold we use the reduced 3D parameter space spanned by $J_y,
J_z, A_x - A_y \equiv 2A$ at fixed values of $J_x=1, A_x + A_y = 0, A_z = 0$.
Here the integrability condition (\ref{Int}) reads
\begin{equation}\label{Intred}
A(1 + J_y^2 - 2J_z^2 - 2A^2) = 0
\end{equation}
and is satisfied on two intersecting 2D surfaces -- the plane $A=0$ and a
hyperboloid. In any plane $A \neq 0$, integrability thus holds on a pair of
hyperbolic curves. Several such lines are shown in Fig.~\ref{fig1}. The two
intersecting straight lines pertain to $A = \pm1/\sqrt{2}$.
%%%%%%%%%%%%%%%%%%%%%%%%%%%%%%%BEGIN-FIGURE%%%%%%%%%%%%%%%%%%%%%%%%%%%%%%%%
%%%%%%%%%%%%%%%%%%%%%%%%%%%%%%%%END-FIGURE%%%%%%%%%%%%%%%%%%%%%%%%%%%%%%%%%

The level-crossing manifolds are lines in $(J_y,J_z,A)$-space, embedded in the
2D integrability manifold (\ref{Intred}). Table~\ref{tab5} gives parametric
representations of the level crossing lines in $(J_y,J_z,A)$-space for four of
the 16 $H_R^\sigma$ blocks with $K=3$.

The dashed curves in Fig.~\ref{fig1} represent projections onto the
$(J_y,J_z)$-plane of the pairs of level crossing lines pertaining to the
invariant blocks $H_{A1A}^3$ and $H_{B1S}^2$ of (\ref{HRsigma}). In
$(J_y,J_z,A)$-space, the two lines of each pair wrap around the integrability
hyperboloid in such a way that one is the reflection image of the other with
respect to the $J_y$-axis. Points of intersection of the level crossing lines
with planes $A=\mathrm{const}$ are marked as full (open) symbols for $A>0$
($A<0$).

%%%%%%%%%%%%%%%%%%%%%%%%%%%%%%BEGIN-TABLE%%%%%%%%%%%%%%%%%%%%%%%%%%%%%%%%%%%%%%%
\begin{table}[htb]
\begin{tabular}{l|ll}
$H_{A1S}^1$ & $\displaystyle J_z = \frac{\pm\sqrt{2}J_y}{\sqrt{1+J_y^2}}$ & 
$\displaystyle A = \frac{\pm(1-J_y^2)}{\sqrt{2(1+J_y^2)}}$ \\ 
%\hline
$H_{A1A}^3$ & $\displaystyle J_z = \frac{\mp\sqrt{2}J_y}{\sqrt{1+J_y^2}}$ & 
$\displaystyle A = \frac{\pm(1-J_y^2)}{\sqrt{2(1+J_y^2)}}$ \\ 
%\hline
$H_{B1S}^2$ & $\displaystyle J_z =
\frac{\pm(1+4J_y+J_y^2)}{\sqrt{10+16J_y+10J_y^2}}$ & 
$\displaystyle A = \frac{\pm 2(1-J_y^2)}{\sqrt{10+16J_y+10J_y^2}}$ \\ 
%\hline
$H_{B1A}^2$ & $\displaystyle J_z =
\frac{\mp(1-4J_y+J_y^2)}{\sqrt{10-16J_y+10J_y^2}}$ & 
$\displaystyle A = \frac{\pm 2(1-J_y^2)}{\sqrt{10-16J_y+10J_y^2}}$
\end{tabular}
\caption[]{Level crossing lines with $\xi>0$ (upper sign) and $\xi<0$ (lower
  sign) in the reduced parameter space $(J_y,J_z,A)$ of four invariant blocks of
  (\ref{HRsigma}) with $K=3$.} 
\label{tab5}
\end{table}
%%%%%%%%%%%%%%%%%%%%%%%%%%%%%END-TABLE%%%%%%%%%%%%%%%%%%%%%%%%%%%%%%%%%%%%%%%%%%

We have investigated the level crossing manifolds for all 16 invariant blocks of
$H_R^\sigma$ with $K=3$ in the reduced parameter space. There exists exactly one
level crossing line wit $\xi>0$ and one with $\xi<0$ in each case. All lines are infinite and different
from each other. Each line crosses the plane $A=0$ at two of the four symmetry
points $(J_y,J_z) = (\pm 1,\pm 1), (\pm 1,\mp 1)$. These are the only points
with $A=0$ where degenerate levels exist. Each level crossing line thus
represents the 1D slice in $(J_y,J_z,A)$-space of the sheet with
$\xi>0$ or $\xi<0$ of one of the 16 4D level crossing manifolds for $K=3$.

%%%%%%%%%%%%%%%%%%%%%%%%%%%%%%%%%%%%%%%%%%%%%%%%%%%%%%%%%%%%%%%%%%%%%%%%%%%%%%%%
%
\subsection{Dimensionality for arbitrary $K$}\label{secIIIe}
%
%%%%%%%%%%%%%%%%%%%%%%%%%%%%%%%%%%%%%%%%%%%%%%%%%%%%%%%%%%%%%%%%%%%%%%%%%%%%%%%%
Higher-dimensional Hamiltonian matrices exist in the two-spin model (\ref{Ham})
as invariant blocks of (\ref{HRsigma}) for $K=6,10,15,\dots$ in 16 different
realizations each. A real symmetric $K\times K$ matrix $B$ has $\frac 12 K(K+1)$
independent elements. On the level-crossing manifold $L$ of dimensionality $d_L$
(to be determined), two or more of the $K$ eigenvalues are degenerate. The
manifold $L$ maps onto a manifold $Z$ of dimensionality $d_Z=d_L-1$, at least
two eigenvalues are zero.

Two vanishing eigenvalues imply that all minors $|m_{ij}|$ of the determinant
$|B|$ are zero, which yields $K^2$ relations among the matrix elements $B_{ij}$
that must be satisfied. Not all relations are independent. The requirement
$|m_{ij}|=|m_{ji}|$ renders $\frac{1}{2}K(K-1)$ relations redundant. For $K>2$
another $K$ relations are redundant because of the condition
$\sum_iB_{ij}(-1)^{i+j}|m_{ij}| = |B| = 0$.\cite{note1} That leaves
$\frac{1}{2}K(K-1)$ independent relations for a guaranteed pair of zero-energy
levels. Consequently, we have $D_Z=K$, i.e $d_L=K+1$. For $K=3$ we thus recover
the results of the explicit calculation, namely a 4D level-crossing manifold in
a 6D space of independent matrix elements. Both dimensionalities are reduced by
one if we impose the condition of zero trace.

In an alternative approach, the matrix $B$ has two vanishing eigenvalues if the
two lowest-order coefficients, $C_0$ and $C_1$, in the characteristic polynomial
\begin{equation}\label{mkpoly}
|B - \lambda| = \sum_{k=1}^K C_k\lambda^k,
\end{equation}
vanish.\cite{note2} They are sums of products of up to $K$ and $K-1$ matrix
elements, respectively. This condition is equivalent to the $\frac{1}{2}K(K-1)$
conditions that all minors $|m_{ij}|$ vanish. The equivalence of the two
alternative criteria alerts us to the fact that the conditions $C_0=C_1=0$ are
compound conditions, each one equivalent to multiple conditions of the kind
$|m_{ij}|=0$.\cite{note3}

In the context of the two-spin model, all matrix elements are functions of six
Hamiltonian parameters. Not all $\frac 12 K(K-1) -1$ relations which
determine the level-crossing manifold are independent any more.  All evidence
suggests that there remain exactly two independent relations, which then
describe a 4D manifold on the 5D integrability surface in 6D parameter space, no
matter what the matrix dimensionality $K$ is.

It is expected that the level crossing manifold of a system with $K$ levels 
$(E_1 \leq E_2 \leq\ldots\leq E_K)$ consists of $K-1$ distinct 4D sheets where
levels $k$ and $k+1$ are degenerate. In the case $K=3$ we have indeed identified
two sheets and labelled them by the sign of the energy parameter $\xi$.

The two independent relations among the Hamiltonian parameters which determine
the 4D level crossing manifolds involve polynomials of degrees $\propto K$.  The
shape of the level-crossing manifolds thus becomes increasingly convoluted as
$K$ grows larger. Any randomly picked path on the integrability manifold will
thus intersect a given 4D sheet of a level-crossing manifold more and more
frequently. As a consequence, the number of level crossing lines in the reduced
parameter space will increase more rapidly than the numbers of levels present.

%%%%%%%%%%%%%%%%%%%%%%%%%%%%%%%%%%%%%%%%%%%%%%%%%%%%%%%%%%%%%%%%%%%%%%%%%%%%%%%%
%
\subsection{Shape for $K=6,10$}\label{secIIIf}
\begin{figure}[ht]
\centerline{\hspace{1mm}\epsfig{file=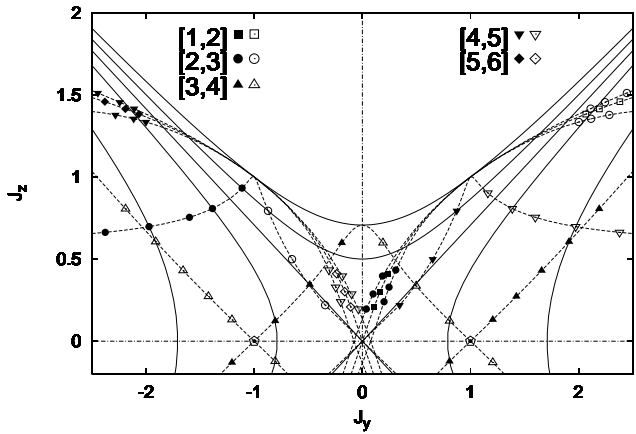,width=6.0cm,angle=-90}}
\caption{The dashed curves are level crossing lines in $(J_y,J_z,A)$-space
  projected onto the $(J_y,J_z)$-plane for the invariant block $H_{A1A}^4$ with
  $K=6$: The solid lines represent the integrability hyperboloid at $A=0, 0.5,
  1/\sqrt{2}, 0.9, 1.4$. The full (open) symbols mark degeneracies between
  levels $k$ and $k+1$ (see legend) at $A>0$ $(A<0)$. Level-crossing lines [3,4]
  in the integrability plane $A=0$ are shown dot-dashed. The pentagons mark the
  positions of two anomalous lines of [3,4] degeneracy perpendicular to the
  $(J_y,J_z)$-plane.}
\label{fig2}
\end{figure}

%
%%%%%%%%%%%%%%%%%%%%%%%%%%%%%%%%%%%%%%%%%%%%%%%%%%%%%%%%%%%%%%%%%%%%%%%%%%%%%%%%
Figure~\ref{fig2} depicts the level crossing manifold for the invariant block
$H_{A1A}^4$ of (\ref{HRsigma}) with $K=6$ levels in the reduced parameter space
$(J_y,J_z,A)$. The representation is similar to that used in Fig.~\ref{fig1} for
$K=3$. The data shown here are mainly the results of a numerical search for
level crossings, but some of the level degeneracies thus identified can be
corroborated analytically. The configuration of level-crossing lines is
reflection-symmetric with respect to the lines $J_y=A=0$ and $J_z=A=0$.

%%%%%%%%%%%%%%%%%%%%%%%%%%%%%%%BEGIN-FIGURE%%%%%%%%%%%%%%%%%%%%%%%%%%%%%%%%
%%%%%%%%%%%%%%%%%%%%%%%%%%%%%%%%END-FIGURE%%%%%%%%%%%%%%%%%%%%%%%%%%%%%%%%%

Among the six levels with energies $E_1 \leq E_2 \leq \ldots\leq E_6$, any
occurrence of a level crossing can be characterized by the position $[k,k+1]$ of
the two degenerate levels in the level sequence.\cite{note4} This label thus
distinguishes five different kinds of level crossings. All level crossing lines
shown in Fig.~\ref{fig2} are labelled accordingly. In the integrability plane
$A=0$, level crossings occur at the four symmetry points $(J_y,J_z) = (\pm 1,\pm
1), (\pm 1,\mp 1)$ as was already the case for $K=3$, and along the two
(dot-dashed) lines $J_y=0$ and $J_z=0$.

On the integrability hyperboloid we have identified ten level crossings lines
(dashed curves) as compared to just two lines for $K=3$. All ten lines are
infinite. Eight of them intersect the integrability plane at the four symmetry
points mentioned previously, where multiple degeneracies occur and are well
understood.\cite{SM90} The intersection points $(J_y=0, J_z=\pm1/\sqrt{2})$ for
the remaining two lines do not involve multiple level degeneracies.

Thus far the structure of the observed level crossing manifold is in full accord
with the scenario outlined in Sec.~\ref{secIIIe}. However, there also exist two
straight lines of level degeneracy oriented perpendicular to the
$(J_y,J_z)$-plane at $(J_y=\pm 1, J_z=0)$. These two level crossing lines are
not confined to the integrability manifold. They involve a degeneracy [3,4] at
energy $E=0$.\cite{Web88,note5} Most important in the context of our study is
the dimensionality of this anomalous level crossing submanifold. Unlike the
other level crossing lines in the reduced parameter space, which are slices of
4D structures in the full 6D parameter space, they remain lower-dimensional.

\begin{figure}[ht]
\centerline{\hspace{1mm}\epsfig{file=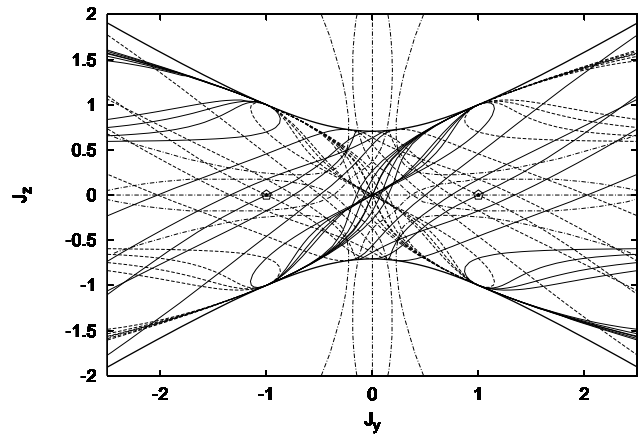,width=6.0cm,angle=-90}}
%\centerline{\hspace{1mm}\epsfig{file=fig3sl.ps,width=6.0cm,angle=-90}}
\caption{Level crossing lines in $(J_y,J_z,A)$-space for the invariant block
  $H_{A1A}^5$ with $K=10$. The solid (dashed) lines are projections onto the
  $(J_y,J_z)$-plane of 30 level crossing lines at $A>0$ $(A<0)$ on the
  integrability hyperboloid. The dot-dashed lines are 10 level crossing lines in
  the integrability plane $A=0$. The thick lines outline  the
  projected hyperboloid. The pentagons mark the positions of two anomalous lines
  of level degeneracy perpendicular to the  $(J_y,J_z)$-plane.}
\label{fig3}
\end{figure}

The data for the invariant block $H_{A1A}^5$ of (\ref{HRsigma}), which has
$K=10$ levels, confirm all the essential features that we have already
identified for the cases $K=3,6$. New features that would necessitate any change
in interpretation have not been observed. Figure~\ref{fig3} shows that the
number of level crossing lines has increased to ten on the integrability plane
$A=0$ (dot-dashed lines) and to 30 on the integrability hyperboloid (solid lines
at $A>0$, dashed lines at $A<0$). As predicted, this increase exceeds the
increase in the number of levels significantly.
%%%%%%%%%%%%%%%%%%%%%%%%%%%%%%%BEGIN-FIGURE%%%%%%%%%%%%%%%%%%%%%%%%%%%%%%%%
%%%%%%%%%%%%%%%%%%%%%%%%%%%%%%%%END-FIGURE%%%%%%%%%%%%%%%%%%%%%%%%%%%%%%%%%

Every level crossing line on the hyperboloid intersects the plane $A=0$ at least
once, either at one of the symmetry points or at the intersection with a level
crossing line in the plane.  The two anomalous level crossing lines observed for
$K=6$ at $(J_y=\pm 1, J_z=0)$ are present again. All level crossing lines except
the anomalous ones represent slices of what must be $K-1=9$ distinct sheets that
make up the 4D level crossing manifold in 6D parameter space. This manifold
remains fully embedded in the 5D classical integrability hypersurface
(\ref{Int}). Only the anomalous submanifold sticks out into the classically
nonintegrable region.

%%%%%%%%%%%%%%%%%%%%%%%%%%%%%%%%%%%%%%%%%%%%%%%%%%%%%%%%%%%%%%%%%%%%%%%%%%%%%%%%
%
\section{Quantum integrability manifold}\label{secIV}
%
%%%%%%%%%%%%%%%%%%%%%%%%%%%%%%%%%%%%%%%%%%%%%%%%%%%%%%%%%%%%%%%%%%%%%%%%%%%%%%%%
The picture that emerges from this study of level degeneracies in a quantum
Hamiltonian system with a nontrivial classical integrability condition may be
summarized as follows: (i) In the 6D parameter space of the two-spin model
(\ref{Ham}), level degeneracies occur predominantly on smooth 4D structures.
(ii) For any given invariant block $H_R^\sigma$ with $K$ levels of the
Hamiltonian matrix (\ref{HRsigma}), this 4D structure consists of $K-1$ sheets,
where each sheet represents one pair $[k,k+1]$ of degenerate levels in the
sequence $E_1 \leq E_2 \leq\ldots\leq E_K$.  (iii) In addition to these 4D level
crossing sheets there also exist lower-D structures in the 6D phase space on
which level degeneracies take place.  (iv) Level degeneracies involving more
than two states, likewise, occur only on lower-D structures. For the most part
they are observed at symmetry points of the Hamiltonian.  (v) All $K-1$ 4D level
crossing sheets pertaining to any invariant block $H_R^\sigma$ are completely
embedded in the 5D hypersurface on which the classical integrability condition
(\ref{Int}) is satisfied. Only lower-D structures of the level crossing manifold
exist elsewhere in parameter space.

These observations are remarkable in the context of the elusive concept of
quantum integrability. One might argue that integrability in the sense of
analytic solvability has no meaning for any matrix $H_R^\sigma$ because
algorithms that diagonalize real symmetric $K \times K$ matrices operate without
any restrictions. The fact is, however, that a universal switch is encoded in
all $H_R^\sigma$ matrices that permits an abundance of level degeneracies on a
smooth 5D hypersurface in 6D parameter space and prohibits them almost
everywhere else, i.e. strictly everywhere else for $K=3$ and everywhere else
except on lower-D submanifolds for $K>3$. As we carry out the analysis for more
and more invariant blocks $H_R^\sigma$, the shape of this 5D hypersurface
emerges with growing definition as the smooth interpolation of an ever
increasing set of 4D level crossing sheets.

There is no a priori reason why the classical integrability condition
(\ref{Int}) should have any such clear-cut bearing on the spectral properties of
low-dimensional irreducible quantum representations of the two-spin model
(\ref{Ham}). On the basis of the correspondence principle, one might surmise
that the 5D classical integrability hypersurface is only relevant quantum
mechanically in an asymptotic sense, i.e. for systems with $\sigma\to\infty$.
The fact is, however, that in some representations with as few as $K=3$ levels
the classical integrability condition results naturally as one of two conditions
that, in combination, guarantee a level degeneracy. Another fact is that (under
mild assumptions) the classical integrability condition (\ref{Int}) can be
reconstructed analytically from the quantum mechanical condition for the
occurrence of level degeneracies within low-$K$ invariant subspaces.

If the level crossing manifolds are described by polynomial equations among the
Hamiltonian parameters as is the case here, then their compatibility with an
integrability condition that is also described by a polynomial is restricted.
B\'ezout's theorem\cite{Bezout} states (effectively) that the maximum number of
independent 4D manifolds which are embedded simultaneously in two different 5D
degree-$n$ polynomial hypersurfaces in projective space is $n^2$. Hence, the 16
independent 4D level-crossing manifolds in 6D parameter space that we have
determined analytically for $K=3$ representations in (\ref{HRsigma}), uniquely
determine the 5D integrability manifold if it is described by a polynomial of
degree less than $\sqrt{16}=4$. For the situation at hand, the classical
integrability manifold is thus the only degree-three polynomial that can
accommodate all 16 level-crossing manifolds for $K=3$. When we add the
polynomial level crossing manifolds for $K=6,10,\ldots$ to the set of embedded
manifolds, the uniqueness of the integrability manifold applies to polynomials
of higher and higher degree.

The relation (\ref{Int}) among the six Hamiltonian parameters is thus no less
relevant for the quantum mechanical properties than it is for the classical
mechanical properties of the two-spin model (\ref{Ham}). It plays the role of a
{\it quantum integrability manifold} as much as it represents the classical
integrability manifold.

The fact that almost all level crossings are confined to this 5D hypersurface in
6D parameter space is compelling indicator that {\it quantum integrability} is a
meaningful concept for systems with few degrees of freedom. However, its essence
has yet to be elucidated. A different indicator of quantum integrability and
nonintegrability, which is based on tracking individual eigenvectors along
closed paths through parameter space is the subject of a study currently in
progress and promises to shed further light on this issue.\cite{SM98}
%%%%%%%%%%%%%%%%%%%%%%%%%%%%%%%%%%%%%%%%%%%%%%%%%%%%%%%%%%%%%%%%%%%%%%%%%%%%%%%%
%
\acknowledgements
%
%%%%%%%%%%%%%%%%%%%%%%%%%%%%%%%%%%%%%%%%%%%%%%%%%%%%%%%%%%%%%%%%%%%%%%%%%%%%%%%%
The work at URI was supported by NSF Grant DMR-93-12252. We are grateful to
J. Stolze for useful comments on the manuscript. 

%%%%%%%%%%%%%%%%%%%%%%%%%%%%%%%%%%%%%%%%%%%%%%%%%%%%%%%%%%%%%%%%%%%%%%%%%%%%%%%%
%

%%%%%%%%%%%%%%%%%%%%%%%%%%%%%%%%%%%%%%%%%%%%%%%%%%%%%%%%%%%%%%%%%%%%%%%%%%%%%%%

\begin{thebibliography}{100}
\bibitem{note1} No further redundancy is caused by the equivalent condition
  $\sum_jB_{ij}(-1)^{i+j}|m_{ij}| = |B| = 0$
\bibitem{note3} The compound nature is already apparent for $K=2$: The two
  conditions $C_0=m_{11}m_{22}-m_{12}^2=0$, $C_1=-m_{11}-m_{22}=0$ are
  equivalent to the three conditions $m_{11}=m_{12}=m_{22}=0$.
\bibitem{note4} The much scarcer events of multiple degeneracy will be
  discussed in passing, but they are of little importance in the context of this
  study.
\bibitem{note5} Weber (Ref.~\onlinecite{Web88}) showed that the energy level
  $E=0$ is $\left( 2 \sigma + 1 \right)$-fold degenerate on the lines $(J_y=\pm
  1, J_z=0)$ in the reduced parameter space, irrespective of whether the
  integrability condition is satisfied or not.  For $\sigma \geq 2$, this number
  $(2 \sigma +1 \geq 5)$ exceeds the number of symmetry classes ($A1A$, $A1S$,
  $B1S$, $B1A$), where $E=0$ does occur. Hence, some eigenvectors within the
  same symmetry class must be degenerate. Weber also showed that for $A \neq 0$
  the degeneracy of $E=0$ is completely removed away from the line $J_y=\pm 1$,
  no matter whether integrability is preserved or not.
\end{thebibliography}
\end{document}